\documentclass{acm_proc_article-sp}[10pt]

\newif\ifpdf
\ifx\pdfoutput\undefined
   \pdffalse     
\else
   \pdfoutput=1  
   \pdftrue
\fi

\usepackage{times}

\ifpdf
\usepackage[pdftex]{graphicx}
\else
\usepackage{graphicx}
\fi

\usepackage{xspace}

\ifpdf
\usepackage[pdftex,bookmarks=false,
            plainpages=false,naturalnames=true,
            colorlinks=true,pdfstartview=FitV,
            linkcolor=blue,citecolor=blue,urlcolor=blue]{hyperref}
\else
\usepackage[dvips]{hyperref}
\fi

\usepackage{amsmath}
\usepackage{amsfonts}
\usepackage{amssymb}
\usepackage{eucal}
\ifpdf
\usepackage[centredisplay]{diagrams}
\else
\usepackage[centredisplay,PostScript=dvips]{diagrams}
\fi

\newcommand{\logrule}[2]{\displaystyle \frac{#1}{#2}}

\newcommand{\isa}{\ensuremath{<}}

\newcommand{\hasa}{\ensuremath{\supset}}
\newcommand{\compose}{\ensuremath{\circ}}
\newcommand{\prcompose}{\ensuremath{\otimes}}
\newcommand{\cocompose}{\ensuremath{\oplus}}
\newcommand{\partinterp}{\ensuremath{\hookrightarrow}}
\newcommand{\partinterps}[2]{\ensuremath{\overset{#1}{\underset{#2}{\partinterp}}}}
\newcommand{\fullinterp}{\ensuremath{\leadsto}}

\newcommand{\realizes}{\ensuremath{:}}
\newcommand{\typefont}[1]{\ensuremath{\mathsf{#1}}}

\newcommand{\logimplies}{\ensuremath{\Rightarrow}}

\newcommand{\kindfont}[1]{\ensuremath{\mathbb{#1}}}

\newcommand{\Kind}{\ensuremath{\kindfont{K}}}

\newarrow{Corresponds} <--->
\newcommand{\kindnamefont}[1]{\textsc{#1}}
\newcommand{\knfont}[1]{\kindnamefont{#1}}
\newcommand{\instancenamefont}[1]{\ensuremath{\textrm{#1}}}
\newcommand{\infont}[1]{\instancenamefont{#1}}

\newcommand{\gamv}{\ensuremath{\Gamma \vdash}}

\newcommand{\parent}{\ensuremath{\isa_{p}}}
\newcommand{\partof}{\ensuremath{\subset_{p}}}
\newcommand{\canon}{\ensuremath{[]}}
\newcommand{\canonform}[1]{\ensuremath{[#1]}}

\newcommand{\equals}{\ensuremath{=}}
\newcommand{\fullequiv}{\ensuremath{\eqcirc}}
\newcommand{\partequiv}{\ensuremath{\lessdot}}

\newcommand{\isakindof}{\ensuremath{\isa_{r}}}
\newcommand{\follows}{\ensuremath{\cdotp}}

\newarrow{Isa} C--->
\newarrow{Hasa} >--->
\newarrow{Equiv} 33333
\newarrow{Realize} |--->
\newarrow{Interpret} ....>
\newarrow{Computes} {}{dash}{}{dash}>

\newarrow{Dashto}{}{dash}{}{dash}>
\newarrow{CorDashto}<{dash}{}{dash}>

\newcommand{\gam}{\ensuremath{\Gamma}}

\newcommand{\wff}{\ensuremath{\vdash \diamond}}

\newcommand{\examplesize}{\footnotesize}
\newcommand{\daysince}{\texttt{days\_\-since\_\-jan1\_\-1970}}

\newtheorem{Definition}{Definition}
\newtheorem{Theorem}{Theorem}

\begin{document}

\conferenceinfo{GCSE/SAIG '02}{2002 Pittsburgh, PA, USA}
\CopyrightYear{2002}

\title{Semantic Component Composition}
\numberofauthors{1}
\author{
\alignauthor Joseph R.~Kiniry\\
       \affaddr{Department of Computer Science}\\
       \affaddr{California Institute of Technology}\\
       \affaddr{Mailstop 256-80}\\
       \affaddr{Pasadena, CA 91125}\\
       \email{kiniry@acm.org}
}

\maketitle

\begin{abstract}
  
  Building complex software systems necessitates the use of component-based
  architectures.  In theory, of the set of components needed for a design,
  only some small portion of them are ``custom''; the rest are reused or
  refactored existing pieces of software.  Unfortunately, this is an
  idealized situation.  Just because two components \emph{should} work
  together does not mean that they \emph{will} work together.
  
  The ``glue'' that holds components together is not just technology.
  The contracts that bind complex systems together \emph{implicitly}
  define more than their explicit type.  These ``\emph{conceptual
  contracts}'' describe essential aspects of extra-system semantics:
  e.g., object models, type systems, data representation, interface
  action semantics, legal and contractual obligations, and more.
  
  Designers and developers spend inordinate amounts of time
  technologically duct-taping systems to fulfill these conceptual
  contracts because system-wide semantics have not been rigorously
  characterized or codified.  This paper describes a formal
  characterization of the problem and discusses an initial
  implementation of the resulting theoretical system.

  
\end{abstract}


                [preprocessors]
                



\section{Introduction} 

Modern software systems are increasingly complicated.  Combine new
technologies, complex architectures, and incredible amounts of legacy
code, and you have systems that no one can even partially comprehend.
As complexity rises, code quality drops and system reliability
suffers.

Building reliable complex systems necessitates the use of excellent
software engineering practices and tools.  But even with this bag of
tricks, the modern software engineer is often overwhelmed with the
technological problems inherent in building such complex systems.
Most projects have to support multiple programming languages and
models, deal with several different architectures (both hardware and
software), have distributed and concurrent subsystems, \emph{and} must
be faster, cheaper, and better than their competitors.

What is worse, while these technological challenges are daunting, it is the
non-technological issues that are often insurmountable.  The widespread
lack of system, design, and component documentation, ignorance about
testing and quality assurance, and lack of communication/knowledge across
teams/companies, make for a very difficult working environment.  The
situation is further compounded by the social problems rampant in
information technology: the Not-Invented-Here syndrome; issues of trust
among companies, teams, and developers; developer education and motivation;
and managerial and business pressures.

\subsection{Semantic Components with\\Conceptual Contracts}

Since developers of reusable constructs (like chunks of code or paragraphs
of documentation) can not say what they mean, there is a
``meaning-mismatch''.  This ``information-impedance'' muddles communication
between collaborators as well as complicates composition between
components.

A new semantic theory, specifically designed to address this and related
problems, can help solve this ``meaning-mismatch''.  If is our belief that
if such a product were widely available and integrated into all aspects of
system development, then our systems would be easier to build, better
documented, and more robust.  Put more plainly, our systems would be
\emph{more correct}.

This paper aims to (1) provide a means by which components can be
semantically (conceptually, formally) described, (2) present an algorithm
which automatically determines when components can inter-operate regardless
of syntax or component model, and (3) describe the means by which the
adapters necessary to compose such components can be automatically
generated.  

A component that is specified, manipulated, and reasoned about using
this (or functionally similar) theory, tools, and techniques is what
we call a \emph{semantic component}.

The theoretical foundation for this work is a new logic for describing
open collaborative reusable assets called \emph{kind theory}, which is
described in full in Kiniry's Ph.D.  dissertation~\cite{Kiniry02-PhD}.

\subsection{Related Work}

Various pieces of research from several communities has similarities to
this work.  At its most simple, straightforward structural subtyping can be
used as a theoretical framework for semantic composition, especially in
strongly typed languages; e.g., Muckelbauer's work in distributed
object-based systems~\cite{Muckelbauer96}.  The problem with this approach
is that only renaming and reordering operations are possible, and little
theoretical infrastructure supports reasoning about the full semantics of
components.

Stronger theoretical formalisms exists for specifying and reasoning about
such compositional systems; for example, the recent work of Molina-Bravo
and Pimentel~\cite{Molina-BravoPimentel02}.  But such work is far from
being practically applicable in the near future as there is no strong
connection with real programming languages, software engineering methods,
and no tool support.

Other expressive formalisms for compositionality exist, primarily in the
category theoretic domain~\cite{FiadeiroMaibaum96}.  This work is starting
to see real application as Fiadeiro has moved into a more industrial role
in promoting this technology.

Work at CMU by Yellin and Strom, inspired by the problems with the Aesop
system~\cite{GarlanAllenOckerbloom95}, has covered this territory before in
the context of object protocols and weak, non-formal semi-automatic adapter
construction~\cite{YellinStrom94, YellinStrom97}.  The work has a very ad
hoc feel, even after the development of a formal model for such
architectures~\cite{AllenGarlan97}.  We plan on using
this model as a motivating example for the evolution of our own design
language, Extended BON~\cite{EBON01}.

In a Java context, Wang et al proposed an service-event model with an
ontology of ports and links, extending the JavaBeans descriptor
model~\cite{WangUngarKlawitter99}.

Other related work comes from the areas of: domain-specific languages,
especially for component specification and re\-use~\cite{Perry99}; the
automatic programming community, e.g, early work by
Barstow~\cite{Barstow79, Barstow85} and Cleveland~\cite{Cleaveland88};
conceptual reuse such as Castano and De
Antonellis~\cite{CastanoDeAntonellis93}; and last but certainly not least,
the software transformation systems community, including the voluminous
works of Biggerstaff and Batory.

Much of this work is reviewed nicely in~\cite{Dusink-vanKatwijk95,
  PartschSteinbruggen83, RichWaters88}.

The primary difference between all of these formalisms and systems and ours
is that our system has a broad, firm theoretic foundation.  That
foundation, kind theory, was specifically designed to reason about reusable
assets in an open collaborative context.  Thus, our work is not tied to a
specific language or realization, integrates the domains of reuse,
knowledge representation, automatic programming, and program
transformation, and does so in a theoretical and systematic framework that
supports global collaboration among many participants.

\section{Semantic Composition} 

Our proposal is, on the surface, relatively simple.  Instead of having only
a syntactic interface to a component, we provide a higher-level semantic
specification.  Additionally, providing this semantic specification does
not entail the need for a developer to learn any new heavyweight
specification language, theory, or tools.

The key to this new solution is the notion of \emph{semantic
  compatibility}~\cite{Kiniry02-SC}.  Components are described
  explicitly and implicitly with lightweight, inline, domain-specific
  documentation extensions called \emph{semantic properties}.  These
  properties have a formal semantics that are specified in kind theory
  and are realized in specific programming languages and systems.

When used in tandem with a computational realization of kind theory (called
a \emph{kind system}), these \emph{semantic components} are composed
automatically through the generation of ``glue'' code, and such
compositions are formally validated.

\section{A Brief Overview of\\Kind Theory} 

\emph{Kind} are classifiers used to describe reusable assets like
program code, components, documentation, specifications,
etc. \emph{Instances} are realizations of kind---actual embodiments of
these classifiers.  For example, the paperback ``The Portrait of the
Artist as a Young Man'' by James Joyce is an \emph{instance} of
\emph{kinds} $\knfont{PaperbackBook}$ and $\knfont{EnglishDocument}$,
(and perhaps others).  We write this as

\begin{gather*}
  \infont{Portrait}\realizes\knfont{PaperbackBook}\cocompose\knfont{EnglishDocument}
\end{gather*}

We use kind theory to specify semantic properties because it provides us
with an excellent model-independent (i.e., it is not bound to some specific
programming language) reuse-centric formalism.

\subsection{Structure}

Kinds are described structurally using our logic in a number of ways
using several core operators.  Classification is covered with the
inheritance operators $\isa$ and $\parent$; structural relationships
are formalized using the inclusion operators $\partof$ and $\hasa$,
equivalence has several forms, $\fullequiv$ and $\partequiv$;
realization, the relationship between instances and kind, is
formalized with the operators $\isakindof$ and $\realizes$;
composition is captured in several forms, $\prcompose$, $\cocompose$,
and $\compose$; and interpretation, the translation of kind to kind or
instances to instances, is realized with the operators $\partinterp$
and $\fullinterp$.

\subsection{Operators}

\subsubsection{Inheritance}

\emph{Inheritance} is a relation that holds between a \emph{child}
kind and one or more \emph{parent} kinds.  The generic notion of
inheritance is that some qualities are transferred by some means from
parent to child.  Inheritance is the relation that is used primarily
to \emph{classify} kind.

The basic symbols that we use for inheritance are $\parent$, $\isa$, and
$\isakindof$ to connote order on kind.  Inheritance is a reflexive,
transitive, asymmetric operation.

Inheritance in the form of classification is one of the most common
structuring notions known in science.  Variants of Aristotle's
\emph{genera} and \emph{species} and the Linnaean method has been used
in various incarnations for hundreds of years to classify organisms
and entities.  Another example of inheritance comes from library
science where researchers have developed several classification
schemas for books: e.g., the Library of Congress and the Dewey Decimal
Systems.  

Some forms of subtyping and inheritance in modern formalisms and
programming languages are sometimes also kinds of inheritance (they
are not when they have nothing to do with semantic conformance).

\subsubsection{Inclusion}

\emph{Inclusion} is a relation that holds between two constructs, the
\emph{whole} and the \emph{part}.  A part is a portion of the whole.  We
use the term \emph{has-a} to connote inclusion.  The basic symbols that we
use for inclusion are $\partof$ and $\hasa$ to connote structural
containment.

Consider this document.  We can hierarchically decompose its entire
physical structure in a variety of different ways.  For example,
syntactically, letters make up words, words make up sentences, sentences
make up paragraphs, etc.  Thus, paragraphs \emph{have} sentences and,
specifically, a given paragraph \emph{has-a} particular sentence.

Inclusion is also a reflexive, transitive, asymmetric operation.

\subsubsection{Equivalence}

\emph{Equivalence} is a relation that holds between constructs that are
the ``same'' in some context.  Two constructs are equivalent if they are
similar for one or more reasons.  The basic symbols that we use for
equivalence are $\equiv$ and $\fullequiv$, and the related $\partequiv$.

Equivalence is a context-sensitive relation.  Consider the two strings
``$Y=2*X$'' and ``$Z=X*2$''.  Textually, they are not equivalent because
they contain different characters.  If interpreted in an algebraic setting,
they are potentially equivalent equations, but it depends upon the specific
algebras in question.  If they both are interpreted in the same algebra,
if that algebra is commutative, and alpha-renaming is a part of the
definition of equivalence, then they are equivalent.  If any of these
conditions fails, or perhaps some other unusual conditions exist (i.e., one
algebra is a modulus group), then the equations are not equivalent.
    
Equivalence is not just something that is applicable to formal
mathematical statements.  Consider a pinball machine and a Sony
PlayStation.  How are these two things equivalent?  Obviously, both are
games that people play for enjoyment, so this classification criterion is
one potential equivalence class.  Another equivalence is that both devices
have buttons, thus some property-based criterion imply equivalence.
Another more subtle equivalence class is the fact that both devices encode
the group $(\mathbb{Z},+)$ because both keep score in some fashion.

\subsubsection{Composition}

\emph{Composition} encapsulates the general notion of taking two or
more constructs and putting them together is some way.  Composition is
a constructive operation---its result is a new thing that has some of
the properties of its constituent pieces.  The semantics of the
specific composition operation used dictates the properties of the new
construct.  We define a generic semantics to which all composition
operations must at least adhere.  The symbols we use for composition
are $\compose$, $\prcompose$, and $\cocompose$.  We read $M \compose
N$ as $M$ \emph{composed with} $N$.

Any constructive operation is composition.  A father putting together
the parts of a bicycle on Christmas Eve is performing a type of
composition; the composition of bicycle parts to make a bike.  The
\texttt{cat} command in UNIX systems is another example of
composition; this time, of two byte streams.

\subsubsection{Realization}

As mentioned earlier, an instance $I$ \emph{realizes} a kind $K$.
\emph{Realization} is the process of stating that a specific thing is
an \emph{instance} of a specific kind.  If $I$ is an instance of a
kind $K$, we say $I$ \emph{realizes} $K$, or $I$ \emph{is an instance
of} $K$, or even $I$ \emph{is of kind} $K$.  We use the symbol
$\realizes$ for realization.  The choice of symbol comes from the
colon character's use in programming languages and type theory.

Realization is the kind theory peer to typing.  When one states ``let
$Z$ be a variable of type $\typefont{integer}$'', a type is being
assigned to the symbol $Z$.  Likewise, when we state $Z \realizes
\knfont{Integer}$ we are \emph{kinding} the symbol $Z$ as having the
kind $\knfont{Integer}$.

\subsubsection{Interpretation}

\emph{Interpretation} is the process of evaluating a construct in some
context, translating it from one form to another.  The symbols that we use
for interpretation are $\partinterp$ and $\fullinterp$.  We read
$K\partinterps{A}{C}L$ as \emph{the partial interpretation kind from kind}
$K$ \emph{to kind} $L$ \emph{by agent} $A$ \emph{in context} $C$.
Likewise, $\fullinterp$ is read \emph{full interpretation}.  When the agent
and context are clear, we simply write $K \fullinterp L$ and read it as
\emph{the full interpretation kind from} $K$ \emph{to} $L$.

An \emph{interpretation} is a computable functional kind that preserves
some type of structure.  We define two sort of interpretation kind.
  
\emph{Full interpretations} are computable functions that preserve all
structure from their domain.  This means that the semantics, that is, the
validity, of all related constructs is maintained across interpretation.
Within kind theory a full interpretation is defined as a \emph{functor} on
a specific category of kind.
  
A \emph{partial interpretation} is a computable function that
preserves some substructure from its domain.  Partial interpretations
are, categorically, \emph{forgetful functors}.
  
Interpretation is a transitive operation.  The identity interpretation
is always defined on all kinds and instances---it is the identity
function, denoted with the term $id$.

Any evaluation process is a type of interpretation.  Reading this document
is one kind of interpretation, evaluating a mathematical expression with
Mathematica is another.  In each case, data is transformed via an agent
(you, the reader, in the former case, and the Mathematica process in the
latter) within a specific context.

\subsubsection{Canonicality}

Associated with every kind $K$ is a full interpretation function
$\canon$, read as ``\emph{canonical}''.  Given a kind (instance) it
returns the \emph{canonical form} of that kind (instance).  The
canonical form of a canonical form is itself.
  
If we have a term of the form $\canonform{k} \equals l$ we call the asset
$l$ the \emph{canonical kind} of $k$.  Likewise, for instances, we use
the term \emph{canonical instance}.  When we do not distinguish kind or
instance, we say \emph{canonical realization} or \emph{canonical
  asset}.
  
Canonicality preserves \emph{all} structure of its domain and, since
the codomain is generative (it is constructed entirely by the
canonicality operation), then it can have \emph{no} new structure.
This means that it is \emph{possible} to define a full interpretation
between any two canonically equivalent assets, but not
\emph{necessary} that such interpretations exist.
  
Thus, canonicality induces an equivalence relation; interpretations do not.

\subsection{Rules}

\begin{table}[htbp]
  \caption{Relevant Rules.}
  \label{tab:Relevant_Rules}

  \begin{equation*}
    \begin{array}{c}
    \text{(Parent Interp*)} \\
    \logrule{\gam,L \fullinterp K, K \partinterp L \vdash \Phi \quad 
             \gamv K \parent L}
            {\gamv L \fullinterp K \follows K \partinterp L \equals
              id_{L}, \Phi} \\
    \\
    \text{(Fully Equiv*)} \\
    \logrule         {\gamv U \fullequiv V}
            {\gamv \canonform{U} \equals \canonform{V}} \\
    \\
    \text{(Partial Equiv*)} \\
    \logrule         {\gamv U \partequiv V}
            {\gamv \canonform{V} \hasa \canonform{U}} \\
    \end{array}
  \end{equation*}
\end{table}

The most important rules of kind theory in the context of this paper
are summarized in Table~\ref{tab:Relevant_Rules}.  The gamma in these
rules is an explicit \emph{context} of sequents (kind theory
sentences).  $\Phi$ is a list of arbitrary sentences, $id_{L}$ is the
identity function on kind $L$, and the asterisk denotes that the rule
is reversible.

The rule (Parent Interp*) states that, if an inheritance relationship
exists between kinds, then two interpretations must also exist: one
that takes the parent to the child, preserving all structure, and a
right adjoint of that map that takes the child to the parent.  This
rule essentially subsumes the related notions of type coercion,
structural type checking, and classification.

The relationship between equivalence and canonicality is elucidated in
the other two rules.  First, two assets, that is, kinds or instances,
are \emph{fully equivalent} if and only if their canonical forms are
identical.  Second, two assets are \emph{partially equivalent} if
their canonical forms are structurally contained---that is, one of
them is part of the other.

\subsection{Semantics}

Semantics are specified in an autoepistemic fashion using what are called
\emph{truth structures}.  Truth structures come in two forms:
\emph{claims} and \emph{beliefs}.

\emph{Claims} are stronger than beliefs.  A mathematically proven statement
that is widely accepted is a claim.  This phrasing is used because, for
example, there are theorems that have a preliminary proof but are not yet
widely recognized as being true.

A statement that is universally accepted, but not necessarily
mathematically proven, is also a claim.  Claims are not necessarily
mathematical formulas.  The statement ``the sun will rise tomorrow'' is
considered by the vast majority of listeners a true and valid statement,
and thus is classified as a claim rather than as a belief.

\emph{Beliefs}, on the other hand, range in surety from \emph{completely
  unsure} to \emph{absolutely convinced}.  No specific metric is defined
for the degree of conviction, the only requirement placed on the associated
belief logic is that the belief degree form a partial order.

\section{Semantic Properties} 

\emph{Semantic properties} are domain-independent documentation constructs
with intuitive formal semantics that are mapped into the semantic domain of
their application.  We use the term ``semantic properties'' because they
are properties (as in property-value pairs) which have formal semantics.

Semantic properties are used as if they were normal semi-structured
documentation.  But, rather than being ignored by compilers and
development environments as comments typically are, they have the
attention of augmented versions of such tools.  Semantic properties
embed a tremendous amount of concise information wherever they are
used without imposing the overhead inherent in the introduction of new
languages and formalisms for similar purposes.

Our current set of semantic properties are listed in
Table~\ref{tab:The_Full_Set_of_Semantic_Properties} in the appendix of this
article, and are specified in full detail in Kiniry's
dissertation~\cite{Kiniry02-PhD}.  To explain their use, we will focus on
particular enabling aspects of kind theory and provide a few examples of
their use.

When used in a particular language, semantic properties are realized using
appropriate domain-specific extensions.  We have integrated their use into
the Java and Eiffel programming languages, as well as in the BON
specification language~\cite{Nerson92, WaldenNerson94}, through a process
that we call \emph{semantic embedding}.

\subsection{Java}

Semantic properties are embedded in Java code using Javadoc-style
comments.  This makes for a simple, parseable syntax.  To give some
flavor to semantic embedding, we'll present a small example of Java
code using semantic properties.  Here is an example of such use, taken
directly from one of our projects that uses semantic
properties~\cite{KiniryIDebug98}.

\scriptsize
\begin{verbatim}
/**
 * Returns a boolean indicating whether any debugging 
 * facilities are turned off for a particular thread.
 *
 * @concurrency GUARDED
 * @require (thread != null) Parameters must be valid.
 * @modifies QUERY
 * @param thread we are checking the debugging condition 
 * of this thread.
 * @return a boolean indicating whether any debugging 
 * facilities are turned off for the specified thread.
 * @review kiniry - Are the isOff() methods necessary at all?
 **/
     
 public synchronized boolean isOff(Thread thread)
 {
   return (!isOn(thread));
 }
\end{verbatim}
\normalsize

The method \texttt{isOff} has a base specification, that of its Java type
signature.  It takes a reference to an object of type \texttt{Thread} and
returns a value of base type \texttt{boolean}.  But more than just its type
signature is used by the Java compiler and runtime.  Additionally, it is a
\texttt{public} method, thus any client can invoke it, and it is
\texttt{synchronized}, thus only a single thread of control can enter it,
or any other synchronized method in the same containing class (called
\texttt{Debug}), at once.  This computational access control is managed by
the Java virtual machine through the use of a monitor object attached to
the \texttt{Debug} class.

Existing tools already use these properties.  Some translate
specifications, primarily in the form of contracts, into run-time test
code.  Reto Kramer's iContract~\cite{Kramer98}, the University of
Oldenburg's Semantic Group's Jass tool, Findler and Felliason's contract
soundness checking tool~\cite{FindlerFelleisen01}, and Kiniry and Cheong's
JPP~\cite{KiniryCheong98} are four such tools.  Other tools translate
specifications into structured documentation.  Javadoc and its relatives
are examples of such tools.

\subsection{Kinding with Semantic Properties}

Semantic properties are used in kinding an instance (e.g., a method, a
class, a type) in the following fashion.  We do not have the space to fully
detail the related theory and algorithm, but a detailed example should
suffice is getting the idea across.  We'll focus exclusively on kinding the
above Java method to this end.

The \emph{kind} of this method contains much more information than its
type.  First, the kind contains everything we have already discussed with
respect to the method's signature.  Additionally, a semantic interpretation
of all of the documentation attached to the method is included.  And, as a
final element, a more complete representation of its type is also
incorporated.

In more detail:
\begin{enumerate}
  
\item All the information that is inherent in the \emph{explicit}
  specification of the method's signature and type are first composed.

  We will call this particular instance $\infont{Debug.isOff}$.
  With respect to classification, this construct is a Java method.  
  This is encoded as
  \begin{gather*}
    \infont{Debug.isOff} \realizes \knfont{JavaMethod}
  \end{gather*}

  We interpret the method's signature as follows:
  \begin{gather*}
  \infont{Debug.isOff.ParameterSet} \partof
  \infont{Debug.isOff} \\
  \infont{Debug.isOff.ReturnType} \partof
  \infont{Debug.isOff} \\
  \end{gather*}
  where
  \begin{gather*}  
  \infont{Debug.isOff.ParameterSet} \realizes
  \knfont{JavaParameterSet} \\
  \infont{Debug.isOff.Parameter0} \partof
  \infont{Debug.isOff.ParameterSet} \\ 
  \infont{Debug.isOff.Parameter0} \realizes 
  \knfont{JavaParameter} \\
  \infont{Debug.isOff.Parameter0Type} \partof
  \infont{Debug.isOff.Parameter0} \\
  \infont{Debug.isOff.Parameter0Name} \partof
  \infont{Debug.isOff.Parameter0} \\
  \infont{Debug.isOff.Parameter0Type} \realizes
  \knfont{JavaType} \\
  \infont{Debug.isOff.Parameter0Type} \equiv
  \infont{java.lang.Thread} \\
  \infont{Debug.isOff.Parameter0Name} \realizes
  \knfont{JavaIdentifier} \\
  \infont{Debug.isOff.Parameter0Name} \equiv
  \infont{thread} \\
  \ldots \text{etc} \ldots
  \end{gather*}
  
  Effectively, we reflectively encode all of the type and signature
  information for the method in a kind theoretic context.  
  
  The structure of the related kinds like $\knfont{JavaParameter}$ encode
  the necessary structure that must be provided by the interpretation, a
  type and a name in this case.  Thus, this part of the process is no
  different than the parsing and typing that goes along with compiling the
  method.
  
\item All the supplementary information embedded via the use of the
  semantic properties and extra-type information is \emph{also} interpreted
  into a kind theoretic context.
  
  For example, the fact that the method is declared as having
  \texttt{GUARDED} concurrency semantics is encoded with the concurrent
  kind:
  \begin{gather*}
    \infont{ConcurrencySemantics0} \partof
    \infont{Debug.isOff} \\
    \infont{ConcurrencySemantics0} \realizes
    \knfont{JavaConcurrencySemantics} \\
    \infont{ConcurrencySemantics0} \equiv
    \infont{GuardedSemantics}
  \end{gather*}
  
  Likewise, the method's visibility, signature concurrency (use of the
  \texttt{synchronized} keyword), side-effects (\emph{modifies}),
  precondition, parameter and return value documentation, and
  meta-information (\emph{review}) are also interpreted.

\item Finally, we interpret a more complete representation of the method's
  type by taking advantage of the domain semantics of refinement.  
  In this example, we attach stronger refinement semantics via the use of
  the specified, explicit, classical contract.  
  
  The existence of the precondition means that, after we interpret such, we
  can: (a) Check that overridden methods properly weaken the precondition
  for behavioral subsumption. (b) Interpret such specification into
  run-time test code---here that entails just a literal insertion of the
  code snippet into the proper place(s) in a rewritten version of the
  method.  (c) Use this behavioral specification as a guard in semantic
  composition (see below).

\end{enumerate}

The \emph{kind} of this method is our formal ``best-effort'' at the
specification of its semantics.  Now, the rules of kind and instance
composition come into play with respect to defining semantic compatibility.

\section{Component Kind} 

Two instances (e.g, objects) are \emph{compatible} if they can
inter-operate in some fashion correctly and in a sound manner.  For
example, the two objects can perform their intended roles in an interactive
manner and the composition of the two objects is as correct as the two
objects when analyzed individually.

An object $O$ is a realization of a semantic component, that is, $O
\realizes \knfont{SemanticComponent}$, if it provides sufficient information
via semantic properties that a kind system can interpret its structure into
a predefined kind.  That is to say, this ``parsing'' step is an
interpretation to some $K$ which is a semantic component.

A $\knfont{SemanticComponent}$ is a kind that contains both
$\knfont{Pro\-vides}$ and a $\knfont{Re\-quires}$ kinds.  Each of
these enclosed kinds specifies a set of kind which the component
exposes, or on which it depends, respectively.  This two substructures
are a generalization of the common \emph{exports} and \emph{imports}
clauses of architecture description and component specification
languages.

If we wish to determine the semantic compatibility of two instances we
use the following formal definition.

\begin{Definition}[Semantic Compatibility]
  Let $I$ and $J$ be two semantic components. That is, 
  \begin{gather*}
    \gam, I \realizes \knfont{SemanticComponent},
    J \realizes \knfont{SemanticComponent} \wff
  \end{gather*}

  Furthermore, $I$ is part of the provisions of an enclosing semantic
  component $C_{P}$, and $J$ is part of the requirements of an enclosing
  semantic component $C_{R}$,
  \begin{gather*}
    \gamv, I \partof P, P \realizes \knfont{Provides}, P \partof C_{P},\\
    J \partof R, R \realizes \knfont{Requires}, R \partof C_{R} \wff
  \end{gather*}

  We say that $P$ and $R$ are \emph{semantically compatible} if an
  interpretation exists that will ``convert'' $R$ into the ontology of $P$,
  that is if there is a $R \fullinterp P$ in, or derivable from, the
  context $\Gamma$.
  
  $I$ and $J$ are \emph{semantically equivalent} if $I \fullequiv J$, of
  course.
\end{Definition}

We can test for the existence of such an interpretation by checking to see
if the canonical forms of $P$ and $R$ structurally contain each other.

\begin{Theorem}[Semantic Compatibility Check]
  \begin{gather*}
    \gam, \canonform{R} \partof \canonform{P} \vdash R \fullinterp P
  \end{gather*}
\end{Theorem}

The proof of this theorem is straightforward.  If $\canonform{R} \partof
\canonform{P}$ then $\canonform{P} \hasa \canonform{R}$ (by definition of
these operators), and thus by the rule (Partial Equiv*), $P \partequiv R$.
By the definition of partial equivalence, when $P \partequiv R$, a full
interpretation exists that takes $P$ to $R$; that is, $P \fullinterp
R$. This full interpretation is exactly the ``conversion'' operator that we
are looking for to guarantee semantic compatibility.

Finally, we say that $I$ and $J$ can be \emph{semantically composed}
if: (a) they are semantically compatible, and (b) their composition is
realizable within their instance domain.

This realization within their instance domain is the ``glue'' code on which
we depend for composition.  We call this construct a \emph{semantic
  bridge}.  Put another way, a \emph{semantic bridge} is a chain of
equivalences between two instances that ensures their contextual base
equivalency.

\section{Examples} 

All the examples below are defined independently of source object
language and ignore the subtle problems of class and type versioning
that are solved in the full system and are not described here.  These
are only illustrative, not prescriptive, examples.

Finally, the ``tight'' coupling demonstrated below is theoretically
equivalent to the more dynamic coupling found in loosely typed
systems.  The same rules and implications hold in such an
architecture.

Some of the following examples will use the following type:
\examplesize
\begin{verbatim}
  Type DateType
      method setDate(day: Integer; 
                     month: Integer; 
                     year: Integer);
      method getDate();
  EndType
\end{verbatim}
\normalsize

Before examining these examples, take note that if two classes are class or
type compatible, they are obviously semantically compatible.

\subsection{Standard Object Semantic Compatibility}

\examplesize
\begin{verbatim}
  Class Date
      method setDate(day: Integer; 
                     month: Integer; 
                     year: Integer);
      method getDate();
  end;

  Class SetDate
      callmethod writeDate(day: Integer; 
                           month: Integer; 
                           year: Integer);
      callmethod readDate();
  end;
\end{verbatim}
\normalsize

The methods tagged with the \texttt{callmethod} keyword are part of
the $\knfont{Requires}$ of the class; those tagged with the
\texttt{method} keyword are part of the $\knfont{Provides}$ part.

These classes are \emph{not} type compatible since their outbound and
inbound interfaces are of two different types (\texttt{Date\-Type} and some
new type; lets call it \texttt{AnotherDateType}).

If the only difference between the methods \texttt{setDate} and
\texttt{writeDate} is \emph{exactly} their syntax, then these classes
\emph{are} semantically compatible.

This mapping exists because both of these classes realize the same
kind, that of $\knfont{Date}$.  We know this fact because at some
point a developer defined this ontology by virtue of a \emph{claim} or
a \emph{belief} on the classes.  Such a truth structure could have
been defined explicitly through the specification of a semantic
property (\emph{realizes}) on the classes, via manipulation in the
development environment, or by direct input with the kind system.

Now, because $\infont{Date} \realizes \knfont{Date}$, then a mapping exists
from each of its parts (e.g., the \texttt{setDate} method) to each of the
parts of the kind (the canonical $\knfont{setDate}$ kind feature).

These maps, when used in composition, define a simple renaming (sort of an
alpha-renaming for instances) for the classes.  We can realize such a
renaming either by directly manipulating the source text (if this is
permissible in the current context), or by generating a simple wrapper
class automatically.  Thus, an adapter which maps calls from
\texttt{writeDate} to \texttt{setDate} and from \texttt{readDate} to
\texttt{getDate} will allow the composition of these two classes to perform
correctly.

\subsection{Extended Object Semantic Compatibility}

The above example is based on a simple syntactic difference between two
classes.  Here is a more complex example.

Consider the following two classes.
\examplesize
\begin{verbatim}
  Class ISODate
      -- requires: year > 1970
      method setDate(year: Integer; 
                     month: Integer;
                     day: Integer);
      method getDate(): ISODate;
  EndClass

  Class SetDate
      -- requires: year > 0
     callmethod setDate(day: Integer; 
                        month: Integer; 
                        year: Integer);
  EndClass
\end{verbatim}
\normalsize

To compose an instance of \texttt{SetDate} with an instance of
\texttt{ISODate}, we have to (a) negotiate the reordering of the parameters of
the \texttt{setDate} method, and (b) check the behavioral conformance of
the composition.  

This reordering is just another simple form of the above
alpha-renaming because the structural operators in kind theory
($\partof$ and $\hasa$) are order-independent.  The behavioral
conformance is verified because $year > 1970 \logimplies year > 0$.

Thus, we can again automatically generate the appropriate code to guarantee
semantic compositionality.

\subsection{Ontological Semantic Compatibility}

Our final example is an example of a solution that would rely more
strongly upon ontology-based semantic information encoded in kind
theory.

Consider the following classes.
\examplesize
\begin{verbatim}
  Class ISODate
      method setDate(year: Integer; 
                     month: Integer; 
                     day: Integer);
      method getDate(): ISODate;
  EndClass

  Class OffsetDate
      method setDate(days_since_jan1_1970: 
                     Integer);
      method getDate(): OffsetDate;
  EndClass
\end{verbatim}
\normalsize

Assume that the parameter \daysince\ was annotated with a reference to a
kind that described \daysince\ meant.  The context of such a description
would necessarily have to have a \emph{ground}--- a common, base
understanding that is universal.

In this case, the \emph{ground} element is the notion of a \emph{day}.  The
relationship between the parameter \daysince\ and the \emph{day} ground
element need be established.

This relationship might be constructed any of a number of correct,
equivalent manners.  For example, a direct interpretation from the triple
\emph{year/month/day} to \daysince\ would suffice.

Or perhaps a more complicated, multistage interpretation would be all
that is available.  For example, the composition of interpretations
from year to month, then month to day, then day to \daysince, would
provide enough information for the generation of a semantic bridge.

This composition of interpretations is automatically discovered and
verified using the semantic compatibility theorem via the rewriting
logic-based component search algorithm of kind theory.

\section{Implementations}

The earliest implementation of this research was done by a student (Roman
Ginis) working with the author in 1998.  We used the Boyer-Moore theorem
prover (Nqthm) to hand-specify the above examples (and more) and prove
semantic compositionality.  Glue code was also written by hand to see what
steps were necessary, in typical structured languages, for realizing the
resulting interpretations.

We have now developed the theoretical infrastructure to completely describe
the semantic components and reason about their composition.  We also have a
realization of the theory in a kind system implemented in SRI's Maude
logical framework~\cite{ClavelEkerEtal96-Principles}. Currently, the
realization of renaming, reordering, and simple interpretations is direct:
it is snippets of Java program code that are explicit realizations of the
corresponding interpretation.

Our next step is to ``lift'' these examples into a general context,
automatically generating code in a variety of contexts, rather than just
using pre-written, parameterized glue code.  

To this end, we have developed a distributed component-based web
architecture called the Jiki~\cite{Jiki98} for use as a test-bed of
semantic compositionality.  The Jiki's components are distributed
JavaBeans, and they interact via a number of technologies including
local and remote method calls, message passing via HTTP and other
protocols, and a tuple-based coordination mechanism based upon Jini's
JavaSpaces.

These components have already been specified with both the Extended
BON~\cite{EBON01} specification language as well as our semantic
properties in their program code.

Thus, our next step is to use this example, complex application with an
existing specification as a testing ground for automatic generation of glue
code for a variety of interface modalities.  In the end, we would like to
be able to use semantic components as a kind of ``\"Uber-IDL'', and
generate glue code that utilizes a large variety of equivalent communication
substrates, both at compile and run-time.

Abstracting that component-based architecture, particularly with a view
toward component quality-of-service, has already been finished.  The result
of which is what we call the \emph{Connector Architecture}, inspired by
Allen and Garlan's similar work~\cite{AllenGarlan94}, and will be
described in a forthcoming paper.

\section{Conclusion} 

We have shown that, using kind theory, we can use the same formalism to
describe software components, reason about their composition, and generate
verifiable ``glue'' code for their composition.  These specifications come
in the form of simple domain-independent annotations to typical program
code, and such annotations are also used for documentation and testing
purposes.

\section{Acknowledgments}

This work was supported under ONR grant JJH1.MURI-1-CORNELL.MURI (via
Cornell University) ``Digital Libraries: Building Interactive Digital
Libraries of Formal Algorithmic Knowledge'' and AFOSR grant
JCD.\-61404-1-AFOSR.\-614040 ``High-Confidence Reconfigurable Distributed
Control''.


\bibliographystyle{plain}

\begin{thebibliography}{10}

\bibitem{AllenGarlan94}
R.~Allen and D.~Garlan.
\newblock Beyond definition use---architectural interconnection.
\newblock {\em ACM SIGPLAN Notices}, 29(8):35--44, 1994.

\bibitem{AllenGarlan97}
Robert Allen and David Garlan.
\newblock A formal basis for architectural connection.
\newblock {\em ACM Transactions on Software Engineering and Methodology},
  6(3):213--249, July 1997.

\bibitem{Barstow79}
D.R. Barstow.
\newblock An experiment in knowledge-based automatic programming.
\newblock {\em Artificial Intelligence}, 12(2):73--119, 1979.

\bibitem{Barstow85}
D.R. Barstow.
\newblock Domain-specific automatic programming.
\newblock {\em IEEE Transactions on Software Engineering}, 11(11):1321--1336,
  November 1985.

\bibitem{CastanoDeAntonellis93}
S.~Castano and V.~{De Antonellis}.
\newblock A constructive approach to reuse of conceptual components.
\newblock In {\em Proceedings of the Second International Workshop on Software
  Reusability}, pages 19--28, 1993.

\bibitem{ClavelEkerEtal96-Principles}
Manuel Clavel, Steven Eker, Patrick Lincoln, and Jos{\'e} Meseguer.
\newblock Principles of maude.
\newblock In {\em In Proc. 1st Intl. Workshop on Rewriting Logic and its
  Applications}, Electronic Notes in Theoretical Computer Science. Elsevier
  Science, Inc., 1996.

\bibitem{Cleaveland88}
J.C. Cleaveland.
\newblock Building application generators.
\newblock {\em IEEE Software}, 5(4):25--33, July 1988.

\bibitem{Dusink-vanKatwijk95}
Liesbeth Dusink and Jan {van Katwijk}.
\newblock Reuse dimensions.
\newblock In {\em Proceedings of SSR '95}, Seattle, WA, 1995.

\bibitem{FiadeiroMaibaum96}
{J.L.} Fiadeiro and T.~Maibaum.
\newblock A mathematical toolbox for the software architect.
\newblock In {\em Proceedings of The Eighth International Workshop on Software
  Specification and Design (IWSSD8 - '96)}, 1996.

\bibitem{FindlerFelleisen01}
R.~Findler and M.~Felleisen.
\newblock Contract soundness for object-oriented languages.
\newblock In {\em Proceedings of Sixteenth International Conference
  Object-Oriented Programming, Systems, Languages, and Applications}, 2001.

\bibitem{GarlanAllenOckerbloom95}
David Garlan, R.~Allen, and John Ockerbloom.
\newblock Architectural mismatch, or, why it's hard to build systems out of
  existing parts.
\newblock In {\em International Conference on Software Engineering}. IEEE
  Computer Society, IEEE Computer Society, May 1995.

\bibitem{KiniryIDebug98}
Joseph~R. Kiniry.
\newblock {IDebug}: An advanced debugging framework for {Java}.
\newblock Technical Report CS-TR-98-16, Department of Computer
  Science,California Institute of Technology, November 1998.

\bibitem{Jiki98}
Joseph~R. Kiniry.
\newblock The {Jiki}: A distributed component-based {Java Wiki}.
\newblock Available via \href{http://www.jiki.org/} {http://www.jiki.org/},
  1998.

\bibitem{EBON01}
Joseph~R. Kiniry.
\newblock The {Extended BON} tool suite.
\newblock \href{http://ebon.sourceforge.net/} {http://ebon.sourceforge.net/},
  2001.

\bibitem{Kiniry02-PhD}
Joseph~R. Kiniry.
\newblock {\em Formalizing Open, Collaborative Reuse with Kind Theory}.
\newblock PhD thesis, California Institute of Technology, 2002.

\bibitem{Kiniry02-SC}
Joseph~R. Kiniry.
\newblock Semantic properties for lightweight specification in knowledgeable
  development environments.
\newblock Submitted for publication, 2002.

\bibitem{KiniryCheong98}
Joseph~R. Kiniry and Elaine Cheong.
\newblock {JPP}: A {Java} pre-processor.
\newblock Technical Report CS-TR-98-15, Department of Computer
  Science,California Institute of Technology, November 1998.

\bibitem{Kramer98}
Reto Kramer.
\newblock {iContract}--­the {Java} design by contract tool.
\newblock In {\em Proceedings of the Twenty-Fourth Conference on the Technology
  of Object-Oriented Languages ({TOOLS} 24)}, volume~26 of {\em TOOLS
  Conference Series}. IEEE Computer Society, 1998.

\bibitem{Molina-BravoPimentel02}
{J.M.} Molina-Bravo and E.~Pimentel.
\newblock Composing programs in a rewriting logic for declarative programming.
\newblock Technical Report LO/0203006v1, Dpto. Lenguajes y Ciencias de la
  Computaci{\'o}n, University of M{\'a}laga, March 2002.

\bibitem{Muckelbauer96}
Patrick~A. Muckelbauer.
\newblock {\em Structural Subtyping in a Distributed Object System}.
\newblock PhD thesis, Purdue University, 1996.

\bibitem{Nerson92}
Jean-Marc Nerson.
\newblock Applying object-oriented analysis and design.
\newblock {\em Communications of the ACM}, 35(9):63--74, September 1992.

\bibitem{PartschSteinbruggen83}
H.~Partsch and R.~Steinbruggen.
\newblock Program transformation systems.
\newblock {\em ACM Computing Surveys}, 15(3):199--236, September 1983.

\bibitem{Perry99}
Dewayne~E. Perry.
\newblock Software evolution and `light' semantics.
\newblock In {\em Proceedings of ICSE '99}, Los Angeles, CA, 1999.

\bibitem{RichWaters88}
C.~Rich and R.C. Waters.
\newblock Automatic programming: Myths and prospects.
\newblock {\em IEEE Computer}, 21(8):40--51, August 1988.

\bibitem{WaldenNerson94}
Kim Wald{\'e}n and Jean-Marc Nerson.
\newblock {\em Seamless Object-Oriented Software Architecture - Analysis and
  Design of Reliable Systems}.
\newblock Prentice-Hall, Inc., 1994.

\bibitem{WangUngarKlawitter99}
Guijun Wang, Liz Ungar, and Dan Klawitter.
\newblock Component assembly for oo distributed systems.
\newblock {\em IEEE Computer}, 32(7):71--78, July 1999.

\bibitem{YellinStrom94}
Daniel~M. Yellin and Robert~E. Strom.
\newblock Interfaces, protocols, and the semi-automatic construction of
  software adaptors.
\newblock In {\em Proceedings of OOPSLA '94}, October 1994.

\bibitem{YellinStrom97}
D.M. Yellin and R.E. Strom.
\newblock Protocol specifications and component adaptors.
\newblock {\em ACM Transactions on Programming Languages and Systems}, 19(2),
  1997.

\end{thebibliography}



\appendix

\section{Semantic Properties Summary}

\begin{table}[htbp]
  \caption{The Full Set of Semantic Properties}
  \label{tab:The_Full_Set_of_Semantic_Properties}
  \begin{center}
    \begin{tabular}{|ccc|}
      \hline
      \textbf{Meta-Information:} & \textbf{Contracts}    & \textbf{Versioning} \\
      author                     & ensure                & version \\
      bon                        & generate              & deprecated \\
      bug                        & invariant             & since \\
      copyright                  & modifies              & \textbf{Documentation} \\
      description                & require               & design \\
      history                    & \textbf{Concurrency}  & equivalent \\
      license                    & concurrency           & example \\
      title                      & \textbf{Usage}        & see \\
      \textbf{Dependencies}      & param                 & \textbf{Miscellaneous} \\
      references                 & return                & guard \\
      use                        & exception             & values \\
      \textbf{Inheritance}       & \textbf{Pending Work} & time-complexity \\
      hides                      & idea                  & space-complexity \\
      overrides                  & review                &  \\
      realizes                   & todo                  &  \\
      \hline
    \end{tabular}
  \end{center}
\end{table}



\end{document}